\documentclass[
    ,final            
  ]
  {aipproc}

\layoutstyle{6x9}

\begin{document}

\title{Baryon semileptonic decays: the Mexican contribution}


\author{Rub\'en Flores-Mendieta}{
  address={Instituto de F{\'\i}sica, UASLP, \'Alvaro Obreg\'on 64, San Luis Potos{\'\i}, S.L.P.\ 78000, Mexico}
}

\author{Alfonso Mart{\'\i}nez}{
  address={ESFM del IPN, Apartado Postal 75-702, M\'exico, D.F.\ 07738, Mexico}
}

\begin{abstract}
We give a detailed account of the techniques to compute radiative corrections in baryon semileptonic decays developed
over the years by Mexican collaborations. We explain how the method works by obtaining
an expression for the Dalitz plot of semileptonic decays of polarized baryons including radiative corrections to
order ${\mathcal O}(\alpha q/\pi M_1)$, where $q$ is the four-momentum transfer and $M_1$ is the mass of the decaying baryon.
From here we compute the totally integrated spin angular asymmetry coefficient of the emitted baryon and compare its value with
other results.
\end{abstract}

\maketitle

\section{Introduction}

Right after Pauli formulated the neutrino postulate, Fermi introduced the field theoretical treatment of the process
$n\to p+e^-+\overline{\nu}_e$ in the early 1930's \cite{pei}. This is the first
known example of a $V$-theory. In the ensuing years other forms of nuclear $\beta$ decays were observed, which prompted Gamow
and Teller to formulate an $A$-theory. The concept of a new class of interactions, the {\it weak interactions}, had just
emerged.

Both the $V$ and $A$ theories were fused into the $V-A$ theory by Sudarshan and Marshak and also independently by
Feynman and Gell-Mann in the late 1950's, motivated by the discovery of parity non-conservation in weak interactions by Lee
and Yang (theoretically) and Wu and Telegdi (experimentally).

In this context the weak interactions are described by an effective Lagrangian
${\cal L}_{\rm eff} (x)= -(G_F/\sqrt{2}) J_\lambda^\dagger(x) J^\lambda (x) + \mbox{h.c.}$,
where $G_F$ is the Fermi coupling constant and the weak current $J_\lambda(x)$ has the $V-A$ structure. $J_\lambda(x)$ can be
separated into weak leptonic $J_\lambda^l(x)$ and weak hadronic $J_\lambda^h(x)$ currents, namely, $J_\lambda(x) =
J_\lambda^l(x) + J_\lambda^h(x)$, where the leptonic current can be written directly in terms of the lepton fields whereas the
hadronic one can be decomposed into parts having definite flavor SU(3) transformation properties and can be written in terms of
quark fields \cite{gk}.

The Lagrangian ${\cal L}_{\rm eff}$ however, faces many problems and cannot be taken as a self-consistent quantum field theory of
weak interactions. Among other aspects, {\it i}) it is not renormalizable; {\it ii}) at high energies it leads to a violation of
unitarity, i.e., it brings in probability non-conservation; and {\it iii}) it has no room for neutral currents.

With the advent of gauge theories, the cornerstone of the theory of weak interactions became the SU(2)$\times$U(1)
Weinberg-Salam theory, which currently possesses a quite impressive experimental evidence \cite{part}.
Further work, both theoretical and experimental, has finally yielded to the
standard model of elementary particles, which embodies our knowledge of the strong and electroweak interactions.

The purpose of this paper is to briefly review the achievements of the past thirty years of theoretical activity in baryon
semileptonic decays (BSD) from the Mexican perspective. We give a detailed account of BSD, focusing on techniques
to the calculation of radiative corrections to observables which have been the major contribution of local research
groups. The paper opens with a historical account of the development of the theoretical approach and continues with an application to the evaluation of radiative corrections to the spin-asymmetry coefficient of the emitted baryon. Numerical results are discussed afterwards. The paper closes with a brief summary and conclusions.

\section{Radiative corrections in BSD: The early years}

In dealing with the radiative corrections (RC) to BSD it is very important to have an
organized systematic approach to deal with the complications that accompany them. These RC depend on an ultraviolet cutoff, on
strong interactions, and on details of the weak interactions, other than the effective $V-A$ theory. In other words, they have a
model-dependent part \cite{sirlin,low}. They also depend on the charge assignments of the decaying and emitted baryons. Their
final form depends on the observed kinematical and angular variables and on certain experimental conditions. Over the years, our
approach to the calculation of these RC has been to advance results which can be established in as-much-as possible once and for
all. This task is considerably biased by the experimental precision attained in given experiments and by the available phase
space in each decay.

A systematic study of the calculation of RC in BSD was initiated back in 1980 \cite{rebeca} following an approach originally
introduced by Sirlin \cite{sirlin} for the electron energy spectrum in neutron decay and later extended to the decay of
polarized neutrons \cite{shann,maya}. An expression for the electron energy spectrum
including radiative corrections was obtained in Ref.~\cite{rebeca}, which was accurate enough to allow experimental analyses to
be performed in a model-independent fashion, provided hard bremsstrahlung photons are experimentally discriminated. At that time,
those results were not directly applicable to the experiments being performed, which had lower statistics and made no provision
to discriminate against hard photons. Later, in the same spirit, the RC to the differential decay rate of polarized neutral and
charged baryons were presented in Ref.~\cite{garcia82}. From that result, obtaining the RC to the decay rates and angular
coefficients was straightforward. This was the first attempt to obtain RC to integrated observables.

A step further was taken in 1987, when in order to improve previous results on the RC to the
electron energy spectrum of baryon $\beta$ decay \cite{garcia82}, a new theoretical expression was obtained, this time
including all terms of order $\alpha q/\pi M_1$ where $q$ is the four-momentum transfer and $M_1$ is the mass of the decaying
baryon \cite{garcia87}. The result is suitable for
model-independent analyses of very-high-statistics experiments, without any experimental constraint on the detection of
hard inner-bremsstrahlung photons. In Ref.~\cite{garcia87} it was shown that the bremsstrahlung contribution to BSD can be computed in a
model-independent way up to terms of first order in $q$, by using the Low theorem \cite{low}.

In the meantime, new high-statistics experiments in BSD \cite{bour} made Dalitz plot (DP) measurements feasible and the application
of RC was necessary. Much of the work had already been advanced in early calculations for the energy spectrum of the charged
lepton and for the decay distribution of polarized baryons \cite{rebeca,maya,garcia82,garcia87}, so the new task was to adapt
the results for the DP. In this respect, an expression for the DP of semileptonic decays of
charged and neutral baryons including RC to order $\alpha$ and neglecting terms of order $\alpha q$ was introduced in
Ref.~\cite{tun89}. The virtual RC presented no new complications. In contrast, the bremsstrahlung RC became rather involved. The
approximation implemented in that work was that the real photons are not observed directly but indirectly
when the energies of the final baryon and charged lepton are found not to satisfy the final three-body overall
energy-momentum conservation. In consequence, a detailed kinematical analysis for determining the integration limits over the
photon variables was mandatory. Besides, the infrared divergence in the bremsstrahlung had to be identified carefully along with
the finite terms that accompany it. A proper choice of variables yielded the integrations feasible. No doubt this paper
marked an important path toward new results. The immediate application was to the computation of RC to the DP
to order $\alpha q/\pi M_1$, both for charged \cite{tun91} and neutral \cite{tun93} decaying baryons.

In summary, from the early years of research we learned that following the analysis of Sirlin of the virtual RC
in neutron beta decay \cite{sirlin} and armed with the theorem of Low for the
bremsstrahlung RC \cite{low} one can show that the model-dependent contributions to both corrections (introducing new form
factors) contribute to orders $(\alpha/\pi)(q/M_1)^n$ with $n=2,3,\ldots$, while orders $n=0,1$ lead to model-independent final
expressions, because their model dependence is absorbed into the already existing form factors. The RC to BSD obtained to these
latter orders are then suitable for
model-independent experimental analyses and are valid to an acceptable degree of precision in the near and intermediate future:
they will be useful in BSD involving heavy quarks.

\section{Radiative corrections to BSD: the recent years}

There are six different charge assignments in BSD, namely, $A^- \rightarrow B^0 (l^- \overline{\nu}_l)$, $A^0 \rightarrow
B^+(l^- \overline{\nu})$, $A^+ \rightarrow B^0 (l^+ \nu_l)$, $A^0 \rightarrow B^- (l^+ \nu_l)$, $A^{++} \rightarrow B^+
(l^+ \nu_l)$, and $A^+ \rightarrow B^{++}(l^- \overline{\nu}_l)$. In Ref.~\cite{rfm02} it was shown that it is not
necessary to calculate each case separately. The final results of the last four cases can be directly obtained from the final
results of the first two cases. This saves a considerable amount of effort, since only the first two cases need be calculated
in detail. Such detail requires the choice of specific kinematic situations. It is the DP that is normally
studied experimentally. However, energy-momentum conservation may allow to discriminate events where photons are emitted
carrying away energy such that the BSD is placed outside the so called three-body region (TBR) of the DP of non-radiative BSD.
The events with those photons belong to what we have referred to as the four-body region (FBR). In addition, when the initial
baryon is polarized the angular correlations between that polarization and the emitted baryon and the emitted charged lepton
involve different RC, and it is not possible to obtain the final results of one correlation from the final results of the other
correlation. Back in the middle 1990's the systematic study of RC to BSD was complete to both orders $n=0$ and $n=1$ for
unpolarized decaying baryons so it was time to tackle new problems by considering the polarization of either the
decaying or emitted baryons. After some work, the analysis was finished to order $n=0$ for polarized decaying baryons, covering
the spin-final baryon momentum and spin-final charged lepton momentum angular correlations \cite{rfm97,rfm00}. Those results were
indeed useful for obtaining theoretical expressions for the angular spin asymmetry coefficients of the emitted baryon and charged
lepton, respectively. Within our approximations and after producing some numbers, our results agreed well with others already
published \cite{toth}.

Nowadays our goal has been extended to cover both the spin-final baryon momentum and
the spin-final charged lepton angular correlations to order $n=1$ of both neutral and negatively charged decaying baryons, restricted to the TBR. The former problem has been already solved \cite{neri2005} whereas the
latter is in progress. A further analysis will take into account the FBR contribution to the RC \cite{rb1} and also the polarization
of the emitted baryon. This will be done in the near future. Let us mention that up to this order our results will be useful in
high-precision experiments involving heavy quarks.

In the next sections we will describe how to apply our methods to get the RC to the TBR of the spin-final baryon momentum
angular correlation to order $n=1$. Basically we borrow some recent results of Ref.~\cite{neri2005} and analyze the case of
a negatively charged decaying baryon in order to illustrate the procedure.

\section{Virtual radiative corrections}

For definiteness, let us consider the BSD
\begin{equation}
A \rightarrow B + l + \overline{\nu}_l, \label{eq:equno}
\end{equation}
and let $A$ denote a negatively charged baryon and $B$ a neutral one, so that $l$ represents a negatively charged lepton and $\overline{\nu}_l$ its accompanying antineutrino. The four momenta
and masses of the particles involved in process~(\ref{eq:equno}) will
be denoted by $p_1=(E_1,{\mathbf p}_1)$, $p_2=(E_2,{\mathbf p}_2)$, $l=(E,{\mathbf l})$, and $p_\nu=(E_\nu^0,{\mathbf p}_\nu)$,
and by $M_1$, $M_2$, $m$, and $m_\nu$, respectively \cite{neri2005}. Further notation and conventions can be found in those references. We organize our results to
explicitly exhibit the angular correlation ${\hat {\mathbf s}_1} \cdot
\hat {\mathbf p}_2$, where $\hat {\mathbf s}_1$ is the spin of $A$. This choice of the kinematical variables yields
the angular spin-asymmetry coefficient of the emitted baryon, denoted hereafter by $\alpha_B$.

The uncorrected transition amplitude ${\mathsf M}_0$ for process (\ref{eq:equno}) is given by
\begin{eqnarray}
{\mathsf M}_0 = \frac{G_V}{\sqrt 2} [\overline{u}_B(p_2) W_\mu(p_1,p_2) u_A(p_1)] [\overline{u}_l(l) O_\mu v_\nu(p_\nu)],
\label{eq:eqdos}
\end{eqnarray}
where $u_A$, $u_B$, $u_l$, and $v_\nu$ are the Dirac spinors of the corresponding particles and 
\begin{eqnarray}
W_\mu (p_1,p_2) & = & f_1(q^2) \gamma_\mu + f_2(q^2) \sigma_{\mu \nu} \frac{q_\nu}{M_1} + f_3(q^2) \frac{q_\mu}{M_1}
\nonumber \\
&  & \mbox{} + \left[g_1(q^2) \gamma_\mu + g_2(q^2) \sigma_{\mu \nu} \frac{q_\nu}{M_1} + g_3(q^2) \frac{q_\mu}{M_1} \right]
\gamma_5. \label{eq:eqtres}
\end{eqnarray}
Here $O_\mu = \gamma_\mu (1+\gamma_5)$, $q\equiv p_1-p_2$ is the four-momentum transfer, and $f_i(q^2)$ and $g_i(q^2)$ are
the conventional vector and axial-vector form factors, respectively, which are assumed to be real in this work.

The observable effects of spin polarization are analyzed by the replacement
\begin{equation}
u_A(p_1) \rightarrow \Sigma(s_1) u_A(p_1) \label{eq:eqcinco}
\end{equation}
where $\Sigma(s_1) = (1-\gamma_5 {\not \! s_1})/2$ is the spin projection operator,
and the polarization four-vector $s_1$ satisfies $s_1 \cdot s_1 = -1$ and $s_1 \cdot p_1 = 0$. In the
center-of-mass frame of $A$, $s_1$ becomes the purely spatial unit vector $\hat {\mathbf s}_1$ which points along the spin
direction.

The virtual RC can be separated into a model-independent part ${\mathsf M}_v$ which is finite and calculable and
into a model-dependent one which contains the effects of the strong interactions and the intermediate vector boson
\cite{rebeca,tun91,tun93}. This
model-dependence can be absorbed into ${\mathsf M}_0$ through the definition of effective form factors, hereafter referred
to as $f_i^\prime$ and $g_i^\prime$. The decay amplitude with virtual radiative corrections
${\mathsf M}_V$ is given by
\begin{equation}
{\mathsf M}_V = {\mathsf M}_0^\prime + {\mathsf M}_v, \label{eq:eqseis}
\end{equation}
where
\begin{equation}
{\mathsf M}_v = \frac{\alpha}{2\pi} \left[ {\mathsf M}_0 \Phi + {\mathsf M}_{p_1} \Phi^\prime \right].
\label{eq:eqsiete}
\end{equation}

The model-independent functions $\Phi$ and $\Phi^\prime$ contain the terms to order ${\mathcal O}(\alpha q/\pi M_1)$
\cite{tun91} and they reduce respectively to $\phi$ and $\phi^\prime$ of Ref.~\cite{tun89},
in the limit of vanishing $\alpha q/\pi M_1$. The second term in Eq.~(\ref{eq:eqsiete}) can also be found in this reference.

At this point we can construct the DP with virtual RC by leaving $E_2$ and $E$ as the relevant variables in the differential
decay rate for process~(\ref{eq:equno}). After making the replacement (\ref{eq:eqcinco}) in (\ref{eq:eqseis}), squaring it,
summing over final spin states, we have
\begin{equation}
\sum_{\rm spins} | {\mathsf M}_V |^2 = \frac12 \sum_{\rm spins} | {\mathsf M}_V^\prime |^2 - \frac12 \sum_{\rm spins} |{\mathsf M}_V^{(s)}|^2. \label{eq:eq11}
\end{equation}

The spin-independent contribution to ${\mathsf M}_V$ in Eq.~(\ref{eq:eq11}), denoted here by ${\mathsf M}_V^\prime$, was
obtained to order ${\mathcal O}(\alpha/\pi)$ in Ref.~\cite{tun89} and later improved to order
${\mathcal O}(\alpha q/\pi M_1)$ in Ref.~\cite{tun91}, so we will borrow the latter result. We now focus
here on the spin-dependent part ${\mathsf M}_V^{(s)}$ to this order of approximation along the lines of
Ref.~\cite{rfm97}, where only terms of order ${\mathcal O}(\alpha/\pi)$ were retained. The
differential decay rate can be written as
\begin{equation}
d\Gamma_V = \frac{dE_2dEd\Omega_2 d\varphi_l}{(2\pi)^5}M_2mm_\nu\left[ \frac12 \sum_{spins} |{\mathsf M}_V^\prime|^2
- \frac12 \sum_{spins} | {\mathsf M}_V^{(s)}|^2 \right]  = d\Gamma_V^\prime - d\Gamma_V^{(s)}, \label{eq:eq12}
\end{equation}
where $d\Gamma_V^\prime$ corresponds to the differential decay rate with virtual RC of unpolarized baryons given by Eq.~(9) of Ref.~\cite{tun91}, except that we have chosen here, without loss of generality, a coordinate frame in the center-of-mass frame of $A$
with the $z$ axis along the three-momentum of the emitted baryon, whereas in Ref.~\cite{tun91} the $z$ axis was chosen along
the three-momentum of the emitted charged lepton. Similarly $d\Gamma_V^{(s)}$ can be obtained by standard techniques.
Thus the decay rate with virtual RC is
\begin{equation}
d\Gamma_V = d\Omega \left\{A_0^\prime + \frac{\alpha}{\pi} (B_1^\prime \Phi + B_1^{\prime \prime} \Phi^\prime) -
{\hat {\mathbf s}_1} \cdot {\hat {\mathbf p}_2} \left[ A_0^{\prime \prime} + \frac{\alpha}{\pi} (B_2^\prime \Phi +
B_2^{\prime \prime} \Phi^\prime) \right] \right\}, \label{eq:eq13}
\end{equation}
where $d\Omega$ is a phase space factor and $A_0^\prime$, $B_1^\prime$, and $B_1^{\prime \prime}$ are given in Ref.~\cite{tun91} whereas $A_0^{\prime \prime}$ can be found in
Ref.~\cite{rfm97}. They all depend on the kinematical variables and are quadratic functions of the form factors.
Equation~(\ref{eq:eq13}) is the differential decay rate with virtual RC to
order ${\mathcal O}(\alpha q/\pi M_1)$. It is model-independent and contains an infrared divergent term in $\Phi$ which
at any rate will be canceled when the bremsstrahlung RC are added.

When including terms of order ${\mathcal O}(\alpha q/\pi M_1)$ the model dependence of the RC shows up.
For the virtual part, one can handle this by introducing effective form factors, $f_i^\prime$ and $g_i^\prime$, in the uncorrected amplitude ${\mathsf M}_0$,
in such a way that \cite{rebeca}
\begin{eqnarray}
\begin{array}{ll}
\displaystyle
f_1^\prime (q^2,p_+\cdot l) = f_1 (q^2) + \frac{\alpha}{\pi} a_1(p_+\cdot l), &
\displaystyle
g_1^\prime (q^2,p_+\cdot l) = g_1 (q^2) + \frac{\alpha}{\pi} b_1(p_+\cdot l), \\ [4mm]
\displaystyle
f_k^\prime (q^2,p_+\cdot l) = f_k (q^2) + \frac{\alpha}{\pi} a_k, &
\displaystyle
g_k^\prime (q^2,p_+\cdot l) = g_k (q^2) + \frac{\alpha}{\pi} b_k, \quad (k = 2,3)
\end{array} \nonumber
\end{eqnarray}
i.e., $f_1^\prime$ and $g_1^\prime$ have a new dependence in the electron and emitted
baryon energies other than the ones in the $q^2$ dependence of the original form factors; this can be seen through $a_1$
and $b_1$, which are functions of the product $p_+\cdot l = (p_1+p_2) \cdot l$. For the remaining form factors, within
our approximations, $a_k$ and $b_k$ $(k = 2,3)$ are constant.

\section{Bremsstrahlung radiative corrections}

In this section we turn to the emission of real photons in the process
\begin{equation}
A \to B +\ell + \overline{\nu}_l + \gamma, \label{eq:eq36}
\end{equation}
where $A$, $B$, $\ell$, and $\overline{\nu}_l$ denote the same particles as in the virtual case and $\gamma$
represents a real photon with four-momentum $k = (\omega,{\mathbf k})$.

\begin{figure}
\caption{\label{fig:fig1} Kinematical region (${\mathsf A}+{\mathsf B})$ as a function of $E$ and $E_2$ of the four-body decay (\ref{eq:eq36}). The region ${\mathsf A}$ corresponds to what is referred to as the three-body region in this work.}
\includegraphics[height=.31\textheight]{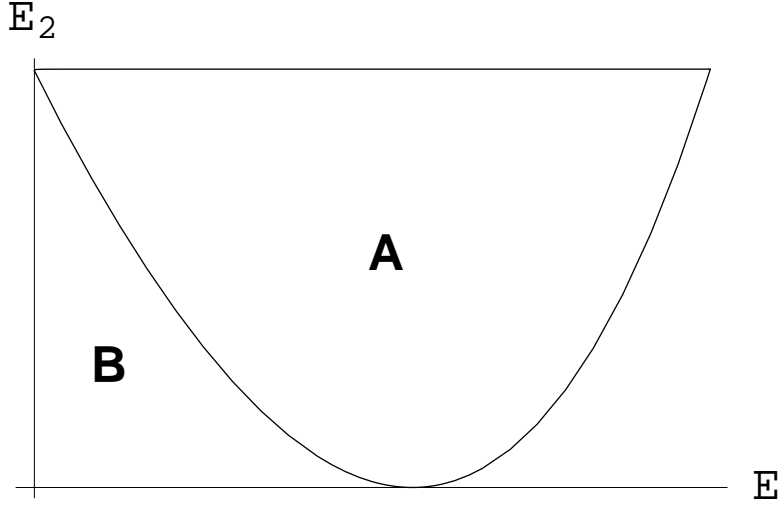}
\end{figure}

The process (\ref{eq:eq36}) itself is strictly speaking a four-body decay whose kinematically allowed region is the joined
area ${\mathsf A} + {\mathsf B}$ depicted in Fig.~\ref{fig:fig1}. The distinction between these two regions has important
physical implications. Region ${\mathsf A}$ is delimited by
\begin{equation}
E_2^{\rm min} \leq E_2 \leq E_2^{\rm max}, \qquad m\leq E \leq E_m \label{eq:eq46}
\end{equation}
where $E_m = (M_1^2-M_2^2+m^2)/2M_1$ whereas region ${\mathsf B}$ is delimited by
\begin{equation}
M_2 \leq E_2 \leq E_2^{\rm min}, \qquad  m\leq E \leq E_b \qquad \label{eq:eq46a}
\end{equation}
with $E_b = [(M_1-M_2)^2+m^2]/2(M_1-M_2)$. Finding an event with energies $E$ and $E_2$ in region ${\mathsf B}$ demands the existence of a fourth particle which must
carry away finite energy and momentum. In contrast, in region ${\mathsf A}$ this fourth particle may or may not do so. Consequently, region ${\mathsf B}$ is exclusively a four-body region whereas region ${\mathsf A}$ is both a three-body and
a four-body region. We will refer to them as the four-body and three-body regions (FBR and TBR), respectively. Our analysis
of the bremsstrahlung RC will consider process (\ref{eq:eq36}) restricted to the TBR.

The starting point will be to obtain the amplitude of process (\ref{eq:eq36}) with the spin effects included, retaining
terms of order ${\mathcal O}(\alpha q/\pi M_1)$ by following the theorem of Low \cite{low}. The amplitude for process (\ref{eq:eq36}), ${\mathsf M}_B$, is given in Ref.~\cite{tun91} and will not be repeated here.

The bremsstrahlung contribution to the DP is obtained from the differential decay rate
\begin{equation}
d\Gamma_B = \frac{M_2mm_\nu}{(2\pi)^8} \frac{d^3p_2}{E_2} \frac{d^3l}{E} \frac{d^3p_\nu}{E_\nu} \frac{d^3k}{2\omega}
\sum_{\rm{spins},\epsilon} | {\mathsf M}_B |^2 \delta^4(p_1-p_2-l-p_\nu-k), \label{eq:eq43}
\end{equation}
where the sum extends over the spins of the final particles and the photon polarization.

In analogy to the virtual RC, the substitution (\ref{eq:eqcinco}) in $\sum |{\mathsf M}_B|^2$ of Eq.~(\ref{eq:eq43}) leads to
\begin{equation}
\sum_{\rm{spins},\epsilon} | {\mathsf M}_B |^2 = \frac12 \sum_{\rm{spins},\epsilon} |{\mathsf M}_B^\prime|^2 - \frac12
\sum_{\rm{spins},\epsilon} |{\mathsf M}_B^{(s)}|^2, \label{eq:eq44}
\end{equation}
and therefore the differential decay rate $d\Gamma_B$ is
\begin{equation}
d\Gamma_B = d\Gamma_B^\prime - d\Gamma_B^{(s)}. \label{eq:eq45}
\end{equation}

Except for minor changes, the quantity $d\Gamma_B^\prime$ in Eq.~(\ref{eq:eq45}) corresponds to the bremsstrahlung
differential decay rate for unpolarized baryons given by Eq.~(42) of Ref.~\cite{tun91}. As for the spin-dependent term,
$d\Gamma_B^{(s)}$, one can find further details about it in Ref.~\cite{neri2005}. The result can be cast into the form
\begin{equation}
d\Gamma_B^{(s)} = \frac{\alpha}{\pi} \, d\Omega \, {\hat {\mathbf s}_1} \cdot {\hat {\mathbf p}}_2 \, [ B_2^\prime \, I_0(E,E_2)
+ C_A], \label{eq:dgbs}
\end{equation}
where $I_0(E,E_2)$ contains the infrared divergent term \cite{rfm97} which cancels the one of its virtual counterpart and $C_A$
in infrared convergent. $C_A$ is presented in two forms in Ref.~\cite{neri2005}. The first one is given in terms of triple
integrals over kinematical variables of the photon and the second one is fully analytic.

The full bremsstrahlung differential decay rate $d\Gamma_B$ is now constructed by
subtracting $d\Gamma_B^{(s)}$ from $d\Gamma_B^\prime$. This
$d\Gamma_B$ is added to $d\Gamma_V$  to obtain
\begin{eqnarray}
d\Gamma = \mbox{} d\Omega \left[A_0^\prime-A_0^{\prime \prime} \, {\hat {\mathbf s}_1} \cdot {\hat {\mathbf p}}_2 + \frac{\alpha}{\pi}
[{\theta_I-\theta_{II}} \, {\hat {\mathbf s}_1} \cdot {\hat {\mathbf p}}_2] \right], \label{eq:dpcom}
\end{eqnarray}
where the functions $\theta_I$ and $\theta_{II}$ can be found in Ref.~\cite{neri2005}.

Expression (\ref{eq:dpcom}) is model-independent and strictly only well-defined in the TBR of the kinematically allowed region. Although this equation is composed of rather lengthy expressions, it has
been organized in such a manner that is easy to use.

\section{Spin-asymmetry coefficient $\alpha_B$}

The DP (\ref{eq:dpcom}) so organized allows the calculation of $\alpha_B$, which is defined as
\begin{equation}
\alpha_B = 2\frac{N^+  -N^-}{N^+ + N^-}.
\end{equation}
Here $N^+$ $[N^-]$ denotes the number of emitted baryons with momenta in the forward [backward] hemisphere with respect to
the polarization of the decaying baryon. Appropriate integration of Eq.~(\ref{eq:dpcom}) leads to
\begin{equation}
\alpha_B = -\frac{B_2+(\alpha/\pi)A_2}{B_1+(\alpha/\pi)A_1}, \label{eq:alfac}
\end{equation}
where
\begin{eqnarray}
\begin{array}{lll}
\displaystyle
B_2 = \int_m^{E_m}\int_{E_2^{\min}}^{E_2^{\max}}A_0^{\prime \prime}dE_2dE, & \qquad &
\displaystyle
A_2 = \int_m^{E_m}\int_{E_2^{\min}}^{E_2^{\max}}\theta_{II}dE_2dE, \\[3mm]
\displaystyle
B_1 = \int_m^{E_m}\int_{E_2^{\min}}^{E_2^{\max}}A_0^\prime dE_2dE, & &
\displaystyle
A_1 = \int_m^{E_m}\int_{E_2^{\min}}^{E_2^{\max}}\theta_IdE_2dE,
\end{array} \nonumber
\end{eqnarray}
and $A_0^{\prime \prime}$, $\theta_{II}$, $A_0^\prime$, and $\theta_I$ are defined in Ref.~\cite{neri2005}.

Equation (\ref{eq:alfac}) is a model-independent analytic expression for $\alpha_B$, including radiative corrections
to order ${\mathcal O}(\alpha q/\pi M_1)$. The uncorrected asymmetry coefficient $\alpha_B^0$ is obtained by dropping the terms proportional
to $\alpha/\pi$ from this equation.

We can evaluate $\alpha_B$ in order to make a comparison with previous works \cite{toth,rfm97}.
We use definite values of the form factors in order to compare under the same
quotations, but this does not mean that our calculation is compromised to any
particular values of them. Therefore we use
$f_1=1.27$, $g_1=0.89$, and $f_2=1.20$ for the decay $\Lambda \rightarrow p e {\overline \nu}$,
$f_1=1$, $g_1=-0.34$, and $f_2=-0.97$ for the decay $\Sigma^- \rightarrow n e {\overline \nu}$, and
$f_1=0$, $g_1=0.60$, and $f_2=1.17$ for the process $\Sigma^- \rightarrow \Lambda e {\overline \nu}$.

First, we can evaluate $\alpha_B (E,E_2)$ [the same Eq.~(\ref{eq:alfac}) without performing the double integrals] in several points of the DP. This evaluation is presented in Table
\ref{table:comth} for the process $\Sigma^- \rightarrow n e {\overline \nu}$, where the first part corresponds to $\alpha_B(E,E_2)$ to order ${\mathcal O}(\alpha/\pi)$ from
Ref.~\cite{rfm97}, the second part is $\alpha_B(E,E_2)$ to order ${\mathcal O}(\alpha q/\pi M_1)$ from this work, and
the last part was computed in Ref.~\cite{toth} and reproduced here for comparison.

\begingroup
\begin{table}
\begin{tabular}{lrrrrrrrrrr}
\hline
$\sigma$ &
\multicolumn{10}{c}{(a)} \\
\hline
 0.8067 &  0.5 &  0.1 &  0.0 &  0.0 &  0.0 &  0.0 &  0.0 &  0.0 &  0.0 &  0.1 \\
 0.8043 & 50.7 &  0.3 &  0.1 &  0.1 &  0.0 &  0.0 &  0.0 &  0.0 &  0.1 &  0.3 \\
 0.8020 &      &  1.2 &  0.3 &  0.2 &  0.1 &  0.1 &  0.1 &  0.0 &  0.1 &      \\
 0.7997 &      &  5.4 &  0.7 &  0.3 &  0.2 &  0.1 &  0.1 &  0.1 &  0.1 &      \\
 0.7974 &      &      &  1.6 &  0.6 &  0.3 &  0.2 &  0.1 &  0.1 &  0.1 &      \\
 0.7951 &      &      &  4.4 &  1.1 &  0.5 &  0.3 &  0.2 &  0.1 &  0.1 &      \\
 0.7928 &      &      & 19.8 &  2.1 &  0.9 &  0.4 &  0.2 &  0.1 &      &      \\
 0.7904 &      &      &      &  5.2 &  1.5 &  0.7 &  0.3 &      &      &      \\
 0.7881 &      &      &      &      &  3.2 &  1.1 &  0.3 &      &      &      \\
 0.7858 &      &      &      &      &  9.8 &  2.2 &  0.2 &      &      &      \\ \\
        & \multicolumn{10}{c}{(b)} \\ \hline
 0.8067 & 0.6 & 0.1 & 0.0 & 0.0 & 0.0 & 0.0 & 0.0 & 0.0 & 0.0 & 0.1 \\
 0.8043 &51.2 & 0.3 & 0.1 & 0.0 & 0.0 & 0.0 & 0.0 & 0.0 & 0.1 & 0.2 \\
 0.8020 &     & 1.2 & 0.3 & 0.1 & 0.1 & 0.0 & 0.1 & 0.1 & 0.1 &     \\
 0.7997 &     & 5.5 & 0.6 & 0.2 & 0.1 & 0.1 & 0.1 & 0.1 & 0.2 &     \\
 0.7974 &     &     & 1.4 & 0.4 & 0.2 & 0.1 & 0.1 & 0.1 & 0.2 &     \\
 0.7951 &     &     & 4.1 & 0.8 & 0.3 & 0.2 & 0.1 & 0.1 & 0.1 &     \\
 0.7928 &     &     &18.5 & 1.7 & 0.6 & 0.3 & 0.2 & 0.1 &     &     \\
 0.7904 &     &     &     & 4.4 & 1.1 & 0.4 & 0.2 & 0.1 &     &     \\
 0.7881 &     &     &     &     & 2.4 & 0.7 & 0.2 &     &     &     \\
 0.7858 &     &     &     &     & 8.3 & 1.4 & 0.1 &     &     &     \\ \\
        & \multicolumn{10}{c}{(c)} \\ \hline
 0.8067 &  0.6 &  0.1 &  0.0 &  0.0 &  0.0 &  0.0 &  0.0 &  0.0 &  0.0 &  0.1 \\
 0.8044 & 50.7 &  0.3 &  0.1 &  0.0 &  0.0 &  0.0 &  0.0 &  0.0 &  0.1 &  0.2 \\
 0.8020 &      &  1.2 &  0.2 &  0.1 &  0.1 &  0.0 &  0.0 &  0.1 &  0.1 &      \\
 0.7997 &      &  5.4 &  0.6 &  0.2 &  0.1 &  0.1 &  0.1 &  0.1 &  0.2 &      \\
 0.7974 &      &      &  1.4 &  0.4 &  0.2 &  0.1 &  0.1 &  0.1 &  0.2 &      \\
 0.7951 &      &      &  4.0 &  0.8 &  0.3 &  0.2 &  0.1 &  0.1 &  0.1 &      \\
 0.7928 &      &      & 18.4 &  1.7 &  0.6 &  0.3 &  0.2 &  0.1 &      &      \\
 0.7904 &      &      &      &  4.4 &  1.0 &  0.4 &  0.2 &  0.1 &      &      \\
 0.7881 &      &      &      & 11.1 &  2.4 &  0.7 &  0.2 &      &      &      \\
 0.7858 &      &      &      &      &  8.2 &  1.4 &  0.1 &      &      &      \\ \\
\hline
$\delta$& 0.05 & 0.15 & 0.25 & 0.35 & 0.45 & 0.55 & 0.65 & 0.75 & 0.85 & 0.95 \\
\hline
\end{tabular}
\caption{\label{table:comth} Percentage $\delta \alpha_B(E,E_2)$ with RC over the TBR in $\Sigma^- \to n e
{\overline \nu}$ decay (a) to order ${\mathcal O}(\alpha/\pi)$ of Ref.~\cite{rfm97}; (b) to order ${\mathcal O}(\alpha q/\pi M_1)$ of this work; and (c) computed in Ref.~\cite{toth}.}
\end{table}
\endgroup

Next, we integrate numerically Eq.~(\ref{eq:alfac}) to obtain $\alpha_B$. This evaluation is displayed in Table \ref{table:cr}, where
the entries of the second column correspond to $\alpha_B^0$, the
following column is the percentage differences defined as $\delta \alpha_B = \alpha_B - \alpha_B^0$,
and the last two columns are reserved to comparisons with Refs.~\cite{rfm97,toth}. There is a very good overall agreement.

\begingroup
\begin{table}
\begin{tabular}{lrrrr}
\hline
Decay & $\alpha_0$ & $\delta \alpha_B$ (this work) & $\delta \alpha_B$ Ref.~\cite{rfm97} & $\delta \alpha_B$ Ref.~\cite{toth} \\
\hline
$\Lambda \rightarrow p e {\overline \nu}$ & $-58.6$ & $-0.09$ & $-0.2$ & $-0.1$ \\
$\Sigma^- \rightarrow n e {\overline \nu}$ & 66.7 & 0.05 & 0.1 & 0.0 \\
$\Sigma^- \rightarrow \Lambda e {\overline \nu}$ & $7.2$ & $0.08$ & & \\
\hline
\end{tabular}
\caption{Values of $\alpha_B$ and comparison with other works.\label{table:cr}
}
\end{table}

\section{Discussion}

We have presented a short review of the situation (past and present) of the achievements in the computation of RC in BSD
developed by Mexican research groups. We have shown how to apply them to the particular case of the spin-asymmetry coefficient
of the emitted baryon and compared with other approaches.

We can claim that the advancement in this topic has been important over the past years so that our understanding on the subject
is now clear. Our approach to compute RC can be used in model-independent analyses for charm-baryon semileptonic decays to a
high degree of precision. Even for semileptonic decays of baryons containing two charm quarks, they provide a good first
approximation. Of course the problem is still open. More work is needed but the approach can be applied straightforwardly.

Finantial support from CONACYT and COFAA-IPN (Mexico) is acknowledged.

\end{document}